\documentclass[fleqn,twoside]{article}
\usepackage{espcrc2}
\usepackage{graphicx}
\usepackage{bm}

\newcommand{\AmS}{{\protect\the\textfont2
  A\kern-.1667em\lower.5ex\hbox{M}\kern-.125emS}}

\title{Moments of nucleon spin-dependent generalized parton
  distributions}

\author{LHPC and SESAM Collaborations:\thanks{This work was supported
    by the U.S.~Department of Energy under contracts DE-AC05-84ER4015,
    DF-FC02-94ER40818 and DE-FG02-91ER40676. Computations were
    performed on the $128$-node Pentium IV cluster at JLab and at
    ORNL, under the auspices of the U.S.~DoE's SciDAC initiative. We
    used MILC gauge field configurations from the Nersc archive.}\\
  W. Schroers\address[MIT]{Center for Theoretical Physics,
    Massachusetts Institute of Technology, Cambridge, MA 02139,
    USA}\thanks{Supported by the Alexander von Humboldt foundation},
  R.C.  Brower\address[UBOS]{Department of Physics, Boston University,
    Boston, MA 02215, USA}, P. Dreher\addressmark[MIT], R.
  Edwards\address[JLAB]{Thomas Jefferson National Accelerator
    Facility, Newport News, VA 23606, USA}, G.
  Fleming\addressmark[JLAB], Ph.
  H{\"a}gler\addressmark[MIT]$^\dagger$, U.M.
  Heller\address[FLU]{American Physical
    Society, One Research Road, Ridge NY 11961-9000, USA}, \\
  Th. Lippert\address[UWUP]{Department of Physics, University of
    Wuppertal, D-42097 Wuppertal, Germany}, J.W.
  Negele\addressmark[MIT], A.V. Pochinsky\addressmark[MIT], D.B.
  Renner\addressmark[MIT], D. Richards\addressmark[JLAB] and K.
  Schilling\addressmark[UWUP]}
       
\begin{document}

\begin{abstract}
  We present a lattice measurement of the first two moments of the
  spin-dependent GPD $\tilde{H}(x,\xi,t)$. From these we obtain the
  axial coupling constant and the second moment of the spin-dependent
  forward parton distribution. The measurements are done in full QCD
  using Wilson fermions. In addition, we also present results from a
  first exploratory study of full QCD using Asqtad sea and domain-wall
  valence fermions.
\end{abstract}
\maketitle
\section{INTRODUCTION}
\label{sec:introduction}
Generalized parton distributions (GPDs) \cite{Muller:1998fv} (see also
\cite{Diehl:2003ny} for a recent review) provide a means of
parametrizing hadronic contributions to both exclusive and inclusive
processes. They reduce in certain limits to form factors and to
(forward) parton distributions. For a review on the nucleon axial
structure see \cite{Bernard:2001rs} and for spin-dependent parton
distributions consult \cite{Goto:1999by}.

GPDs depend on three independent kinematic variables and are therefore
far more difficult to extract from experiments than forward parton
distributions. Lattice simulations provide a general,
model-independent way to compute their moments directly. First results
for spin-independent GPDs have been presented in
\cite{Gockeler:2003jf}. These papers, however, concentrate on rather
large quark masses. It is imperative to extend these studies down into
the chiral regime.

In this talk, we will present a first study of spin-dependent GPDs,
both with Wilson fermions at large quark masses and with staggered sea
and domain-wall valence fermions at intermediate quark masses.

\section{PARAMETRIZATION}
\label{sec:parametrization}
Spin-dependent GPDs are specified by $\tilde{E}^{\mbox{\tiny
    f}}(x,\xi,t)$ and $\tilde{H}^{\mbox{\tiny f}}(x,\xi,t)$, defined
via
\begin{eqnarray}
  & \!\!\!\! \bar{p}^+\int\frac{dz^-}{2\pi} e^{\mbox{\tiny
  i}\bar{p}^+z^-} \langle p'\vert \bar{\psi}^{\mbox{\tiny f}}(-z^-/2)
  \gamma_5 \gamma^+ \psi^{\mbox{\tiny f}}(z^-/2)\vert p\rangle &
  \nonumber \\
  & \!\!\! = \tilde{H}^{\mbox{\tiny f}}(x,\xi,t) \langle\!\langle
  \gamma_5\gamma^+\rangle\!\rangle - \tilde{E}^{\mbox{\tiny
  f}}(x,\xi,t)\frac{\Delta^+}{2m}
  \langle\!\langle\gamma_5\rangle\!\rangle \,. &
  \label{eq:spin-dep-def}
\end{eqnarray}
The upper index f denotes the quark flavor, $x$ is the average
longitudinal momentum fraction of the struck quark, and $\xi$ the
longitudinal momentum transfer. The total invariant momentum transfer
squared is given by $t\equiv\Delta^2$, with the four-momentum transfer
$\Delta=p'-p$. The average hadron momentum is denoted by
$\bar{p}=(p'+p)/2$. We also use the short-hand notation
$\langle\!\langle \Gamma\rangle\!\rangle=\bar{u}(p')\Gamma u(p)$.

By taking moments with respect to $x$, we end up with a tower of local
matrix elements of the form
\begin{equation}
  \langle p'\vert \bar{\psi}^{\mbox{\tiny f}}
  \gamma^{\lbrace\mu_1} \gamma_5 \mbox{i}D^{\mu_2}\cdots\mbox{i}
  D^{\mu_n\rbrace} \psi^{\mbox{\tiny f}}\vert p\rangle\,.
  \label{eq:me-def}
\end{equation}
\begin{table}[th]
  \begin{tabular}[b]{*{3}{c|}c}
    \hline
    \multicolumn{4}{c}{SESAM $\Omega=16^3\times 32$} \\
    \multicolumn{4}{c}{$\beta=5.6, N_{\mbox{\tiny f}}=2$ Wilson}
    \\ \hline
    Num & \multicolumn{2}{c|}{$\kappa_{\mbox{\tiny
      sea}}=\kappa_{\mbox{\tiny val}}$} & $a^{-1}/\mbox{GeV}$ 
    \\ \hline
    $197$ & \multicolumn{2}{c|}{$0.1560$} & $2.01(1)$ \\
    $205$ & \multicolumn{2}{c|}{$0.1565$} & $2.08(2)$ \\
    $194$ & \multicolumn{2}{c|}{$0.1570$} & $2.16(3)$ \\ \hline
    \multicolumn{4}{c}{MILC $\Omega=20^3\times 64$} \\
    \multicolumn{4}{c}{$\beta=6.85, N_{\mbox{\tiny f}}=3$ Asqtad}
    \\ \hline
    Num & $am_{\mbox{\tiny s}}$ & $am_{\mbox{\tiny u+d}}$ &
    $a^{-1}/\mbox{GeV}$ \\ \hline
    $105$ & $0.05$ & $0.05$ & $1.507(6)$ \\ \hline
    \multicolumn{4}{c}{$\beta=6.76, N_{\mbox{\tiny f}}=2+1$ Asqtad}
    \\ \hline
    $105$ & $0.05$ & $0.01$ & $1.464(5)$ \\ \hline
  \end{tabular}
  \caption{Working points and simulation parameters.}
  \label{tab:latt-pars}
\end{table}
These matrix elements can then be computed by a lattice simulation.
The parametrization of these matrix elements follows from their
Lorentz-structure in the continuum and is expressed in terms of the
generalized form factors (GFFs) $\tilde{A}_{ni}^{\mbox{\tiny f}}$ and
$\tilde{B}_{ni}^{\mbox{\tiny f}}$. For example, for $n=2$:
\begin{eqnarray}
  & \!\!\!\!\! \langle p'\vert\bar{\psi}^{\mbox{\tiny f}}
  \gamma^{\lbrace\mu}
  \gamma_5 \mbox{i}D^{\nu\rbrace}\psi^{\mbox{\tiny f}} \vert p\rangle
  & \nonumber \\ & \!\!\!\!\! = \tilde{A}^{\mbox{\tiny
  f}}_{20}(t)\langle\!\langle 
  \gamma^{\lbrace\mu}\gamma_5\rangle\!\rangle \bar{p}^{\nu\rbrace} +
  \tilde{B}^{\mbox{\tiny f}}_{20}(t) \frac{\mbox{i}}{2m}
  \langle\!\langle\gamma_5\rangle\!\rangle \bar{p}^{\lbrace\mu}
  \Delta^{\nu\rbrace} &
\end{eqnarray}
The moments of $\tilde{E}^{\mbox{\tiny f}}(x,\xi,t)$ and
$\tilde{H}^{\mbox{\tiny f}}(x,\xi,t)$ are polynomials in $\xi^2$ with
$\tilde{A}^{\mbox{\tiny f}}_{ni}(t)$ and $\tilde{B}^{\mbox{\tiny
    f}}_{ni}(t)$ as coefficients,
\begin{eqnarray}
  \int dx\, x^{n-1}\, \tilde{H}^{\mbox{\tiny f}}(x,\xi,t) &=&
  \sum_{i=0}^{n/2} 
  (2\xi)^{2i} \tilde{A}^{\mbox{\tiny f}}_{n(2i)}(t)\,, \nonumber \\
  \int dx\, x^{n-1}\, \tilde{E}^{\mbox{\tiny f}}(x,\xi,t) &=&
  \sum_{i=0}^{n/2} 
  (2\xi)^{2i} \tilde{B}^{\mbox{\tiny f}}_{n(2i)}(t)\,.
  \label{eq:sd-decomp}
\end{eqnarray}
The reconstruction of the GPDs is therefore possible by an inverse
Mellin transform.

\section{LATTICE SIMULATION}
\label{sec:lattice-simulation}
We use five samples of unquenched gauge field data in our simulations.
The parameters of the lattices are presented in
tab.~\ref{tab:latt-pars}. As valence quarks we use Wilson fermions on
the SESAM lattices and domain wall fermions with a height of $M=1.7$
and $L_5=16$ on the MILC lattices. In the latter case we also use
HYP-smearing \cite{Hasenfratz:2001hp} with $\alpha_1=0.75,
\alpha_2=0.6,$ and $\alpha_3=0.3$.
\begin{figure}[th]
  \includegraphics[clip=true,scale=0.28]{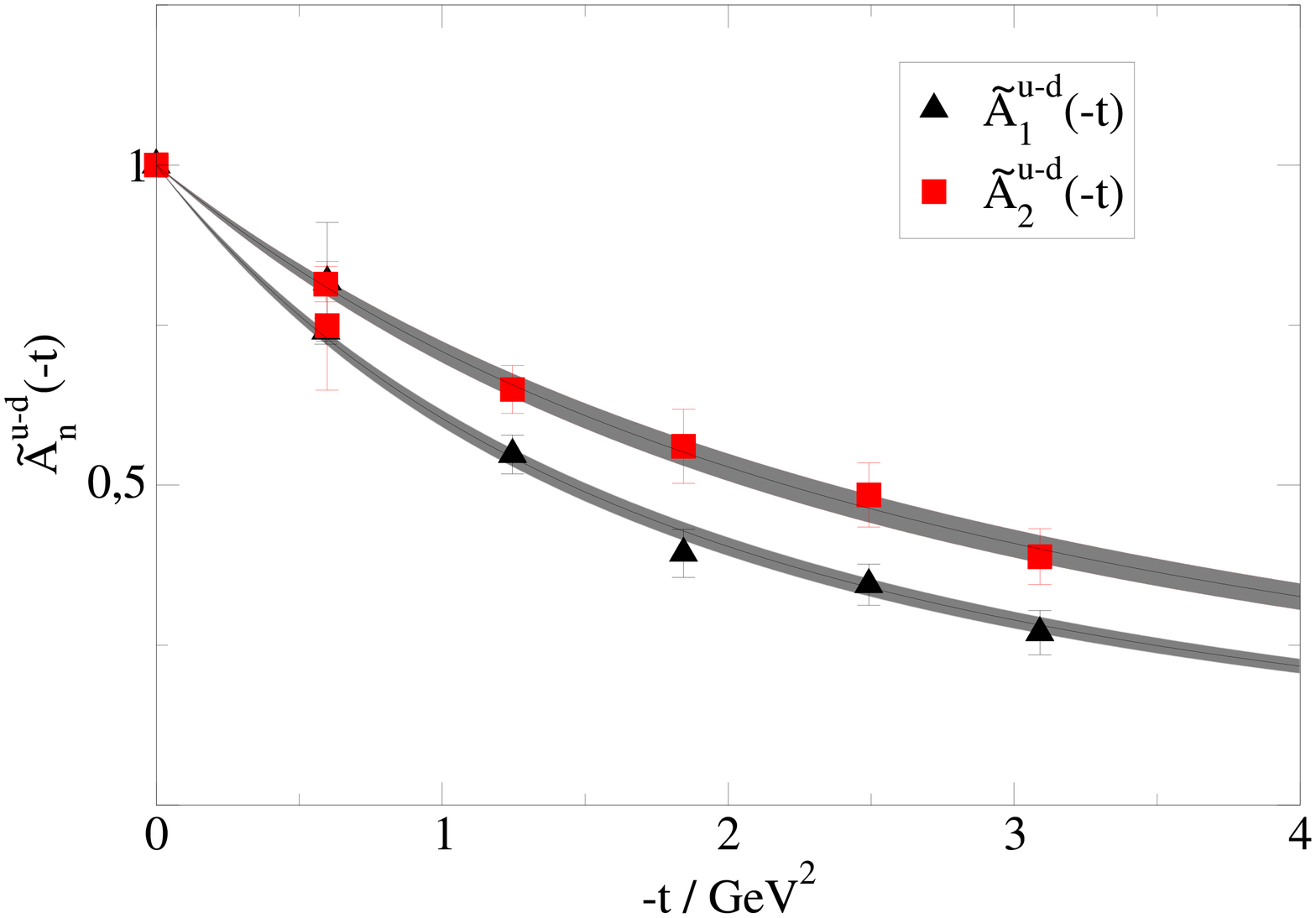}
  \caption{GFFs $\tilde{A}^{\mbox{\tiny u-d}}_{n0}(t)$ with $n=1,2$ for
    $\beta=5.6$, $\kappa_{\mbox{\tiny sea}}=\kappa_{\mbox{\tiny
        val}}=0.1560$. The form factors have been normalized to one at
    $t=0$ and fitted by a dipole form.}
  \label{fig:atilde-n}
\end{figure}
The domain-wall masses have been adjusted to keep the pseudoscalar
lattice mass in the region of the lowest corresponding staggered one.
For the Wilson fermion renormalization constants we use the
perturbative one-loop results quoted in \cite{Dolgov:2002zm}. The
renormalization constants for the domain-wall case are not yet
calculated, so we use the tree-level value. Hence, our results are
preliminary.
\begin{figure}[tb]
  \includegraphics[clip=true,scale=0.28]{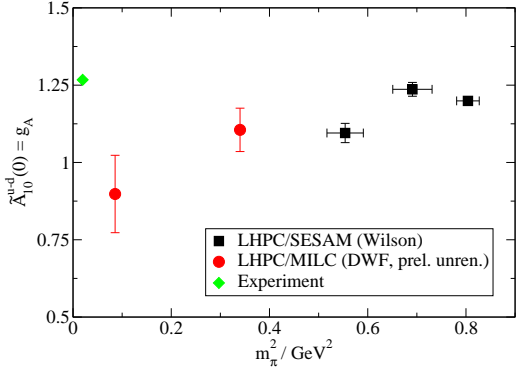}
  \caption{Axial coupling constant $g_A = \tilde{A}^{\mbox{\tiny
        u-d}}_{10}(0)$.}
  \label{fig:axialcoup}
\end{figure}
\begin{figure}[tb]
  \includegraphics[clip=true,scale=0.28]{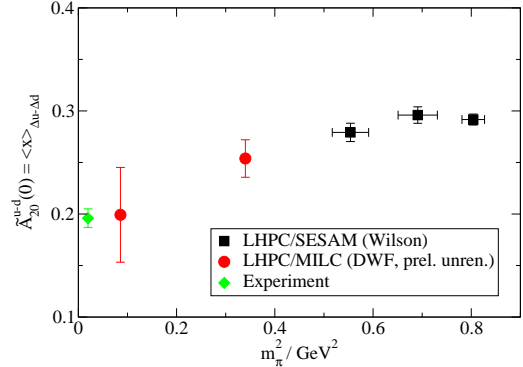}
  \caption{First moment of the spin-dependent parton distribution
    $\langle x\rangle_{\Delta u-\Delta d} = \tilde{A}^{\mbox{\tiny
        u-d}}_{20}(0)$.}
  \label{fig:atilde-forw}
\end{figure}

We concentrate on the quark flavor combination u-d since the resulting
matrix elements are free from disconnected contributions. The GFF
$\tilde{A}^{\mbox{\tiny u-d}}_{10}(t)$ corresponds to the axial form
factor, while $\tilde{A}^{\mbox{\tiny u-d}}_{20}(t)$ is the first GFF
which is not directly accessible experimentally. Both GFFs are plotted
with normalization $\tilde{A}^{\mbox{\tiny u-d}}_{n0}(0)=1$ for the
heaviest quark mass in fig.~\ref{fig:atilde-n}. The curves provide
dipole fits to the data points with the error bands representing one
standard error. It is apparent that the dependencies on the parameters
$x$ and $t$ of $\tilde{H}^{\mbox{\tiny u-d}}(x,\xi,t)$ do not
factorize, a result that is very similar to the spin-independent case
\cite{Negele:2003ar}. However, the difference between the two moments
appears to be smaller in the spin-dependent case.

The axial coupling as a function of the quark mass is plotted in
fig.~\ref{fig:axialcoup}. One should note, however, that this quantity
is highly sensitive to finite-volume effects \cite{Hemmert:2003pr}. At
least at the lightest mass, a couple of simulations at larger lattice
volumes need to be performed to achieve a conclusive result for the
chiral behavior.

The first moment of the forward parton distribution
$\tilde{A}^{\mbox{\tiny u-d}}_{20}(0)$ is displayed in
fig.~\ref{fig:atilde-forw}. Although the measured values decrease in
the chiral regime toward the experimental value, this result needs to
be corroborated with better statistics.

\section{CONCLUSIONS}
\label{sec:conclusions}
In this talk we have presented first results on spin-dependent
generalized parton distributions. In the forward case we have
presented preliminary results for light quark masses which eventually
should allow us to bridge the gap to the chiral regime.

While the axial coupling may be contaminated by substantial
finite-size effects the first moment $\tilde{A}^{\mbox{\tiny
    u-d}}_{20}(0)$ of the spin-dependent GPD appears to be compatible
with experiment in the chiral regime.

\end{document}